\documentclass[sigconf=true, nonacm=true]{acmart}
\title{Benchmarking the Status of Default Pseudorandom Number Generators in Common Programming Languages}
\author {
    Nils van den Honert, 
    Diederick Vermetten, 
    Anna V. Kononova
}
\affiliation {
    Leiden Institute for Advanced Computer Science\\
    nilsvdhonert@gmail.com, d.l.vermetten@liacs.leidenuniv.nl, a.kononova@liacs.leidenuniv.nl \\
}

\begin{document}
\begin{abstract}
	The ever-increasing need for random numbers is clear in many areas of computer science, from neural networks to optimization. As such, most common programming language provide easy access to Pseudorandom Number Generators. However, these generators are not all made equal, and empirical verification has previously shown some to be flawed in key ways. Because of the constant changes in programming languages, we perform the same empirical benchmarking using large batteries of statistcal tests on a wide array of PRNGs, and identify that while some languages have improved significantly over the years, there are still cases where the default PRNG fails to deliver sufficiently random results.

\end{abstract}
\maketitle

\section{Introduction}

In modern computer science, random numbers are a core dependency of a wide array of different application types. From neural networks to iterative optimization heuristics, huge amounts of random numbers need to be generated. As such, Pseudorandom Number Generators (PRNGs) have been introduced in nearly all common programming languages. However, the implementation details of these generators can be challenging to verify, and faulty implementations have the potential to lead to bias within the applications using the affected PRNG~\cite{DBLP:journals/soco/BirdEF20}. Thus, rigorous benchmarking and analysis of the PRNGs has to be performed to ensure the generators keep up with the ever increasing demand for random numbers. 

While there has been quite a lot of research into the properties of different types of PRNG, this research is often only theoretical in nature. From an empirical standpoint, several independent tests have been performed over the years, but their results are no longer up-to-date. In this paper, we use these established benchmarking techniques to analyze the current versions of default PRNGs within a wide set of common programming languages.

\section{Testing PRNGs}
Several different toolboxes for the evaluation of PRNGs have been introduced. These suites generally consist of a large collection of statistical tests for assessing different properties of random numbers. For our purposes, we consider only those which check the uniform distribution $\mathcal{U}(0,1)$. We identify the three most popular benchmarking suites for this application: 
\begin{itemize}
    \item DIEHARDER  test suite~\cite{Brown2011RobertPage}, which is an evolution of the DIEHARD test suite.
    \item NIST test suite~\cite{Rukhin2010AApplications}.
    \item TestU01 suite~\cite{LEcuyer2007TestU01:Generators}.
\end{itemize}
Among these, the TestU01 test suite was found to be the most rigorous, and is thus selected as the testset to apply to the PRNGs we consider. We will focus on the so-called default generator withing each programming language, which means it is described within the documentation of the language as a way to generate random numbers.
 

Within the TestU01 suite, there are several batteries of statistical tests with differing levels of stringency. These batteries are SmallCrush, Crush, and BigCrush, and they differ not only in the intensity of the testing procedures, but also in the types of tests included. Globally speaking, the included tests can be split into four categories:
\begin{itemize}
    \item tests measuring global uniformity by assessing a single stream of numbers,
    \item tests on sub-sequences of fixed length,
    \item tests on grouped vectors of numbers,
    \item tests checking the time until specific events occur in a stream of numbers.
\end{itemize}


\section{Results}
To test the PRNGs in a wide set of different programming languages, we need to make use of a language-agnostic process of transfering the generated random numbers to the TestU01 process, which is written as a C-library. While this could be achieved by storing numbers in files, this would require a significant storage overhead as the largest test batteries can require us to $2^{38}$ random numbers. Instead, we make use of the named pipeline architecture, which creates a connection between the output of one program (the generator) and the input of another (the TestU01 library). The code used for this process is made available on GitHub\footnote{\url{github.com/Nils134/PRNG}}.



The results of the statistical tests in the TestU01 batteries are judged based on their resulting p-values. We can identify two thresholds of p-value which are of interest~\cite{LEcuyer2007TestU01:Generators}: suspect values ($p<10^{-10}$) and failures ($p<10^{-300}$). Given the large numbers of samples which can be used by the statistical tests, suspect p-values are an indication of issues with the generator, but do not definitively show a failure. 



To ensure the testing performed in this paper is reproducible, we will use the same seed to initialize all PRNGs. We verified that changing the seed has little impact on the number of reported failing p-values, as was also shown in~\cite{LEcuyer2007TestU01:Generators}. 




For our testing, we select a set of 17 programming languages.
For each of these, we apply the different batteries of TestU01 and report their results in Table~\ref{tab:results}. Note that the more stringent tests are not run when the previous tests show complete failure rates. 

\begin{table}[ht]
    \centering
    \small
    \caption{Number of suspect (subscript) and failure p-values resulting from the different test batteries of TestU01.}
    \begin{tabular}{p{2.2cm}|p{0.7cm}|c|p{0.4cm}p{0.4cm}p{0.4cm}p{0.4cm}}
        Language  &\rotatebox{90}{Type} &  \rotatebox{90}{Version} &  \rotatebox{90}{SmallCrush} & \rotatebox{90}{Crush} & \rotatebox{90}{BigCrush} & \rotatebox{90}{DIEHARD}   \\ 
        \hline
        Julia & MT & 1.0.4 & 0 & 1 &  1 &  0 \\ 
        Matlab & MT & R2021a & 0 &  1 &  1 &  0 \\ 
        NumPy & MT& 3.6.9 & 0 & $2_{1}$ & $2_{1}$ & 0  \\ 
        Octave & MT & 6.2.0 & 0 & 0 & 1 & 0 \\
        Python & MT  & 3.6.9 & 0 & 1 & 1 &$1_{1}$ \\
        
        PHP  & MT& 7.2 & 0 & $2_{1}$  & 2 & 1 \\ 
        R & MT& 4.0.5 & 0 & 0 & 2 & 1  \\ 
        
        Rust & MT & 1.53.0 &  0 & 0 & 2 & 0 \\ 
        Visual Basic & MT & 16.9 & 0 & 1 & 2 & 0 \\ 
        \hline
        Bash & LCG & 4.4.20 & 15 & -- & -- & 126 \\
        C  & LCG& 7.5.0 & 15 & -- & -- & 126 \\ 
        C++ & LCG & 7.5.0 & 15 & -- & -- & 126 \\
        Java  & LCG& 8 & 1 & $12_{3}$ & $22_{1}$ & $8_{4}$\\ 
        \hline
        C\# & Lfib & 8.0 & $2_{1}$ & $11_{2}$ & 11 & 2 \\
        \hline
        Fortran & XSR & 2018 & 0 & 0 & 0 & 0 \\ 
        \hline
        Mathematica & LSFR& 12.3 & 0 & 0 &  0 &  0   \\ 
        \hline
        Javascript & N/A & 16 & 0 &  0 &  1 &  0   \\ 
        \hline
        \multicolumn{3}{r|}{Tests in the suite} & 15 & 144 & 161 & \multicolumn{1}{c}{126}       
    \end{tabular}
    \label{tab:results}
\end{table}


We see that some generators, such as Bash, fail even on the SmallCrush battery. This is caused by the fact that the implementation for these languages relies on generating random integers with a maximum of 32 bits and normalizing these results to obtain $\mathcal{U}(0,1)$. This results in random numbers which are not granular enough to pass the required tests.

Two additional PRNGs were tested, specific for the field of heuristic optimization. One is the generator used to initialize the random transformation of the functions in the state-of-the-art heuristic benchmarking suite BBOB-platform, while the other is used in an alternative benchmarking suite within the IOHexperimenter tool\footnote{numbbo.github.io, https://iohprofiler.github.io/} (PBO). The former fails on 1, 9 and 9 tests SmallCrush, Crush and BigCrush respectively, while the latter -- in 11 out of 15 tests of Crush. 

\begin{figure}
    \centering
    \includegraphics[scale=0.58,trim=0.8mm 2.1mm 2.5mm 3.1mm,clip]{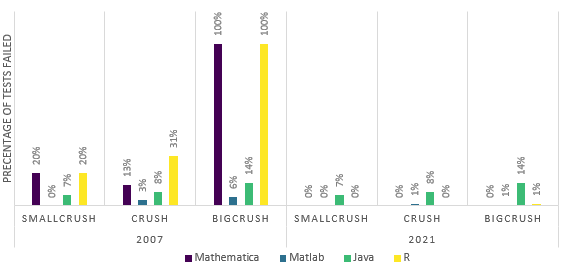}
    \caption{Evolution of languages that have been tested by L'Ecuyer and Simard with TestU01 \cite{LEcuyer2007TestU01:Generators} }
    \label{fig:percentages}
\end{figure}

In Figure~\ref{fig:percentages}, we zoom in on a few of these languages for which their PRNGs have been tested in the past \cite{LEcuyer2007TestU01:Generators}, comparing the number of failures on the four batteries of tests. This figure clearly shows that the updates which these languages, with the exception of Java, have done to their generators have been highly beneficial, almost completely eliminating the previous failures.


\section{Conclusions}

From the results shown in Table~\ref{tab:results}, we can see that there are languages which have implemented PRNGs which pass all test in even the strictest batteries they were exposed to. While these generators might be more complex than the standard MT or LCG versions, their benefit is clear. 

While most applications will not be significantly impacted by failures of the tests in BigCrush, it is still useful to know the properties of the exact generator used, so applications where the random numbers are critical can adopt a different generator if needed.




\section{Future Work}
As we have shown in this paper, the state of PRNGs within popular programming languages is not static. Their implementation might be updated, new packages can be introduced and even new languages adopted. As such, performing robust benchmarking of  PRNGs within the context of even-expanding requirements is needed to ensure the unbiasedness of programs which make use of these generators. 

\bibliographystyle{ACM-Reference-Format}
\bibliography{aaai22.bib}

\end{document}